\documentclass[pra,preprint,showpacs,aps,amsmath,amssymb]{revtex4-1}
\usepackage{epsf}
\usepackage{color}
\usepackage{dcolumn}
\usepackage{bm}
\usepackage[dvipsnames]{xcolor}
\usepackage{revsymb}
\everymath{\displaystyle}

\begin{document} 

\title{Leading correction to the relativistic Foldy-Wouthuysen Hamiltonian}

\author{Alexander J. Silenko}
\email{alsilenko@mail.ru} \affiliation{Bogoliubov Laboratory of Theoretical Physics, Joint Institute for Nuclear Research, Dubna 141980, Russia}


\begin {abstract}
For Dirac particles interacting with external fields, we use the exact operator of the Foldy-Wouthuysen transformation obtained by Eriksen and rigorously derive a leading correction in the weak-field approximation to the known relativistic Foldy-Wouthuysen Hamiltonian. For this purpose, we carry out the operator extraction of a square root in the Eriksen operator. The derived correction is important for the scattering of relativistic particles. Since the description of this scattering by a relativistic wave equation of the second order is more convenient, we determine a general connection between relativistic wave equations of the first and second orders. For Dirac particles, the relativistic wave equation of the second order is obtained with a correction similar to that to the Foldy-Wouthuysen Hamiltonian.
\end{abstract}
\maketitle

\section{Introduction}\label{Introduction}

The seminal transformation of relativistic Hamiltonians to the Schr\"{o}dinger-like form proposed by Foldy and Wouthuysen \cite{FW} is now widely used not only in physics but also in quantum chemistry.
The Foldy-Wouthuysen (FW) representation has unique properties. In this representation, the Hamiltonian and all operators are even, i.e., block-diagonal (diagonal in two spinors or spinor-like wave functions). Relations between
operators (including the spin) in the FW representation are similar to those
between the respective classical quantities. The form of quantum-mechanical operators for relativistic particles
in external fields is the same as in the nonrelativistic quantum theory. In particular, the position (Newton-Wigner)
operator \cite{NW} and the momentum one are equal to $\bm r$ and $\bm p=-i\hbar\nabla$, respectively (see Refs. \cite{Reply2019,PRAFW} for more details). The polarization operator for
spin-1/2 particles is defined by the Dirac matrix $\bm \Pi$ and is expressed by much more cumbersome
formulas in other representations (see Refs. \cite{JMP,PRAFW,FW}). A great advantage of the FW representation is the simple
form of operators corresponding to classical observables.
The passage to the classical limit usually reduces to a replacement of the operators in quantum-mechanical Hamiltonians and equations of motion
with the corresponding classical quantities. The possibility of such a replacement, explicitly
or implicitly used in practically all works devoted to the FW transformation, has been
rigorously proved in Ref. \cite{JINRLett12}. The approach applied in this paper to the relativistic FW Hamiltonian is similar to the approach used by Wentzel, Kramers, and Brillouin in the nonrelativistic case and the conditions of the Wentzel-Kramers-Brillouin approximation should be satisfied. The relativistic operator equations of motion have been obtained \cite{JINRLett12} with the general Hamilton equations. Such a similarity between the quantum-mechanical and classical Hamiltonians and equations of motion is perfectly confirmed by numerous examples (see Ref. \cite{PRAFW} and references therein). The comparison of commutators of fundamental quantum-mechanical operators in different representations and Poisson brackets of corresponding classical variables shows (see Ref. \cite{PRAFW}) that counterparts of the classical variables are the corresponding FW (but not Dirac) operators. The FW representation extends the Schr\"{o}dinger one to  relativistic energies. The probabilistic interpretation of wave functions lost in the Dirac representation is restored in the FW one \cite{Reply2019,PRAFW}. Thanks to these properties, the FW representation provides the best possibility of obtaining a meaningful classical limit of relativistic quantum mechanics \cite{FW,dVFor,CMcK}. 

While Eriksen obtained the exact formula \cite{E} for the FW transformation operator (see Sec. \ref{Previously}), explicit expressions for FW Hamiltonians describing particles in external fields cannot be obtained except in a few particular cases. In the present study, we use the Eriksen formula for a derivation of a leading correction to the previously obtained \cite{PRA2015} \emph{relativistic} FW Hamiltonian. The Eriksen formula is applied for such a goal for the first time. We describe a single particle with a spin 1/2 in stationary and \emph{nonstationary} external fields. The consideration of the nonstationary case has some specific features (see Ref. \cite{PRAnonstat}).

We use the system of units $\hbar=1,~c=1$ but include $\hbar$ and $c$
explicitly when that inclusion
clarifies the problem. Our denotations of the Dirac matrices follow Ref. \cite{BLP}.

The paper is organized as follows. Previously obtained results related to the topic of the present study are considered in Sec. \ref{Previously}. In particular, we analyze main methods of the FW transformation.
The appropriate operator extraction of a square root in the operator of the exact FW transformation is fulfilled in Sec. \ref{Extraction}. In Sec. \ref{FWH}, we derive the relativistic FW Hamiltonian with the leading correction in the weak-field approximation. The corresponding relativistic wave equation of the second order is found in the general form in Sec. \ref{SecondOrderE}. A Dirac particle in an electrostatic field and a spin-1/2 (Dirac-Pauli \cite{Pauli}) one with an anomalous magnetic moment in nonstationary electric and magnetic fields are considered as examples in Sec. \ref{Example}. The obtained results are discussed and summarized in Sec. \ref{Discussion}.

\section{Previously obtained results} \label{Previously}

The passage to the classical limit usually reduces to a replacement of the operators in quantum-mechanical Hamiltonians and equations of motion
with the corresponding classical quantities. The possibility of such a replacement, explicitly
or implicitly used in practically all works devoted to the FW transformation, has been
rigorously proved in Ref. \cite{JINRLett12}. The probabilistic interpretation of wave functions lost in the Dirac representation is restored in the FW one \cite{PRAFW}. Thanks to these
properties, the FW representation provides
the best possibility of obtaining a meaningful classical limit
of relativistic quantum mechanics \cite{FW,CMcK}.

In the general case, the initial Hamiltonian can be written in the form 
\begin{equation} {\cal H}=\beta {\cal M}+{\cal E}+{\cal
O},\qquad\beta{\cal M}={\cal M}\beta,\qquad\beta{\cal E}={\cal E}\beta,
\qquad\beta{\cal O}=-{\cal O}\beta. \label{eq3G} \end{equation} The even operators ${\cal M}$ and ${\cal E}$ and the odd operator ${\cal O}$ are
diagonal and off-diagonal in two spinors, respectively. This
equation is applicable for a particle with any spin if the number
of components of the corresponding wave function is equal to
$2(2s + 1)$, where $s$ is the spin quantum number. 

In the present study, we focus attention on spin-1/2 particles and present the initial Hamiltonian as follows (${\cal M}=mc^2$):
\begin{equation} {\cal H}=\beta mc^2+{\cal E}+{\cal
O},\qquad\beta{\cal E}={\cal E}\beta,
\qquad\beta{\cal O}=-{\cal O}\beta. \label{eq3Dirac} \end{equation} 

In the general case, a transformation to a new representation
described by the wave function $\Psi'$ is performed with the
unitary operator $U$:
\begin{equation} \Psi'=U\Psi=\exp{(iS)}\Psi. \label{eqUiS} \end{equation}

This transformation involves not only the Hamiltonian operator but also the $-i\frac{\partial}{\partial
t}$ one. As a result, the Hamiltonian operator in the new representation takes the form
\begin{equation} {\cal H}'=U\left({\cal H}-i\hbar\frac{\partial}{\partial
t}\right)U^{-1}+ i\hbar\frac{\partial}{\partial t}  \label{taeq2} \end{equation} or
\begin{equation} {\cal H}'=U{\cal H}U^{-1}-i\hbar U\frac{
\partial U^{-1}}{\partial t}. \label{eq2}
\end{equation}

Equation (\ref{taeq2}) can be written in the form
\begin{equation} {\cal H}'-i\hbar\frac{\partial}{\partial
t}=U\left({\cal H}-i\hbar\frac{\partial}{\partial
t}\right)U^{-1}=U\left(\beta{\cal M}+{\cal E}+{\cal
O}-i\hbar\frac{\partial}{\partial
t}\right)U^{-1}.
\label{taeq3}
\end{equation}
This equation presents a rather important property of the FW transformation for a particle in nonstationary (time-dependent) fields \cite{PRA2016}. 
Transformations of two even operators, ${\cal E}$ and
$-i\hbar\frac{\partial}{\partial t}$, are very similar. As a
result, the FW Hamiltonian (except for terms without commutators)
contains these operators only in the combination ${\cal F}={\cal
E}-i\hbar\frac{\partial}{\partial t}$. Therefore, a transition
from a stationary to a nonstationary case can be performed by a
replacement of ${\cal E}$ with ${\cal F}$ in all terms containing
commutators \cite{PRA2016}.

The operator of the FW transformation can be presented in the exponential form:
\begin{equation} U_{FW}=\exp{(i\mathfrak{S}_{FW})}.\label{Vvetott} \end{equation}
The condition eliminating the ambiguity in the definition of this operator has been proposed by
Eriksen \cite{E} and has been substantiated by Eriksen and Korlsrud
\cite{erik} (see also Ref. \cite{PRA2016} for more details). The transformation remains \emph{unique} if the
operator $\mathfrak{S}_{FW}$ in Eq. (\ref{Vvetott}) is \emph{odd},
\begin{equation} \beta\mathfrak{S}_{FW}=-\mathfrak{S}_{FW}\beta,
\label{VveEfrt} \end{equation} and Hermitian
($\beta$-pseudo-Hermitian for bosons \cite{PRA2016}). Further explanation of the
Eriksen method is given in Ref. \cite{PRA2016}.

Condition (\ref{VveEfrt}) is equivalent to \cite{E,erik}
\begin{equation} \beta U_{FW}=U^\dag_{FW}\beta.\label{Erikcon} \end{equation}

Thus, the FW transformation operator should satisfy Eq.
(\ref{Erikcon}) and should perform the transformation in one step.
Eriksen \cite{E} has found an operator possessing these
properties in the stationary case. To determine its explicit form, one can introduce the sign
operator $\lambda={\cal H}/({\cal H}^2)^{1/2}$ and can use the fact
that the operator $1+\beta\lambda$ cancels either lower or upper
spinor for positive and negative energy states, respectively. The numerator and denominator of the operator $\lambda$ commute. It
is easy to see that \cite{E}
\begin{equation}\lambda^2=1, \quad [\beta\lambda,\lambda\beta]=0, \quad [\beta,(\beta\lambda+\lambda\beta)]=0.\label{eq3X3}
\end{equation}
Therefore, the operator of the exact FW transformation has the form \cite{E}
\begin{equation}
U_{E}=U_{FW}=\frac{1+\beta\lambda}{\sqrt{2+\beta\lambda+\lambda\beta}},
~~~ \lambda=\frac{{\cal H}}{({\cal H}^2)^{1/2}}, \label{eqXXI}
\end{equation}
where the operator $({\cal H}^2)^{1/2}$ should be even if ${\cal E}=0$. In this case, squaring Eq. (\ref{eq3Dirac}) and extracting the square root show that $({\cal H}^2)^{1/2}=(m^2+{\cal O}^2)^{1/2}$. To unambiguously define the square root, this relation should be complemented by the condition that the square root of the unit matrix ${\cal I}$ is equal to the unit matrix \cite{JMP}. The initial Hamiltonian operator, ${\cal H}$, is arbitrary. The numerator and denominator of the operator $U_{E}$ commute. The even operator $\beta\lambda+\lambda\beta$ acting on the wave function with a single nonzero
spinor cannot make another spinor be nonzero.

The equivalent form of the operator $U_{E}$ \cite{JMPcond} shows that it is properly unitary ($\beta$-pseudounitary for bosons):
\begin{equation}
U_{E}=\frac{1+\beta\lambda}{\sqrt{(1+\beta\lambda)^\dag(1+\beta\lambda)}}.
\label{JMP2009}
\end{equation}

The Eriksen operator (\ref{eqXXI}) can be used for a particle with
any spin. In this case, the initial Hamiltonian is given by Eq.
(\ref{eq3G}). An exact exponential FW transformation operator has been found in Ref. \cite{PRAExpO}.

Evidently, the Eriksen method gives the right relativistic FW Hamiltonian for a free Dirac particle \cite{FW}: 
\begin{equation}
{\cal H}_{FW}=\beta\epsilon, \qquad \epsilon=\sqrt{m^2+{\cal O}^2}=\sqrt{m^2+\bm p^2}.
 \label{eqX}
\end{equation} The substantiation of the Eriksen method in the more
general case of ${\cal E}\neq0$ and $[{\cal O},{\cal E}]=0$  has
been fulfilled in Ref. \cite{Valid}. In this case, ${\cal H}_{FW}$ is also exact \cite{Valid}.

There are semi-relativistic and relativistic
methods of the FW transformation. We use the term
``semi-relativistic'' for methods \cite{FW,Steph,Reuse,ultrafast}
using an expansion of a derived block-diagonal Hamiltonian in a series in powers of the momentum and potential divided by the mass [$p/(mc)$ and $V/(mc^2)$
]. For relativistic methods, the FW Hamiltonian is expanded in a series in powers of the potential divided by the total kinetic energy ($V/\sqrt{m^2c^4+c^2p^2}$). Sometimes only leading terms are calculated. The zeroth-order Hamiltonians are the Schr\"{o}dinger and FW Hamiltonians of a free particle for the semi-relativistic and relativistic methods, respectively. 

The first semi-relativistic method proposed
by Foldy and Wouthuysen \cite{FW} is iterative. The result of all iterations (with unitary transformation operators) can be asymptotically presented as the single unitary operator $U_{res}=\exp{(i\mathfrak{S}_{res})}$ \cite{erik,dVFor,JMPcond,PRA2016}. This operator and, therefore, the corresponding method lead to the FW representation if and only if $\mathfrak{S}_{res}$ and $U_{res}$ satisfy Eqs. (\ref{VveEfrt}) and (\ref{Erikcon}). Paradoxically, the original method by Foldy and Wouthuysen
\cite{FW} does not satisfy this requirement (the operator $\mathfrak{S}_{res}$ contains even terms and is not odd). Therefore, it \emph{does not lead} to the FW representation \cite{erik,dVFor,JMPcond,PRA2016}. A possibility to correct the original method \cite{FW} has been shown in Ref. \cite{PRA2016}. In this paper (see also Refs. \cite{E,erik,dVFor,JMPcond}), main distinctive features of the FW transformation have also been considered.

Thus, we should differentiate the transformation to the FW representation and the transformation via iterations which has been originally proposed by Foldy and Wouthuysen \cite{FW} and does not lead to this representation. Some other iterative semi-relativistic FW transformation methods satisfy the Eriksen conditions (\ref{VveEfrt}) and (\ref{Erikcon}) at any step of transformations and lead to the FW representation.

Some of relativistic FW transformation methods developed in quantum mechanics are based on unitary
transformations \cite{JMP,FizElem,PRA,PRA2015,PengReiher} and a number of
these methods uses different approaches \cite{relativistic}. All FW transformation methods applied in
quantum chemistry of heavy atoms are relativistic. These methods mainly 
follow the Douglas-Kroll-Hess approach \cite{DouglasKroll,Hess} (consisting in an expansion in a series in powers of $V/\sqrt{m^2c^4+c^2p^2}$) but often use different
transformation operators
\cite{QuantChem,ReiherWolf,ReiherWolfNext,local}. As a rule, relativistic FW transformation methods applied in quantum chemistry meet the Eriksen conditions. 
We refer to the books \cite{Dyall,ReiherWolfBook} for more details.

We should mention that all semi-relativistic methods fail to determine atomic wave functions near a nucleus. When $p/(mc)\ge1$ or $|V|/(mc^2)\ge1$, any series diverges. This is not appropriate for quantum chemistry of heavy atoms ($Z\sim10^2$) even when wave functions obtained with semi-relativistic methods are asymptotically correct far from a nucleus. The impossibility to determine the wave functions near the nucleus prevents the use of the semi-relativistic methods for a calculation of magnetic moments and other parameters of heavy atoms. However, such methods are appropriate for a calculation of the corresponding parameters of \emph{light} atoms if resulting transformation operators satisfy the Eriksen conditions (\ref{VveEfrt}) and (\ref{Erikcon}). For such atoms, the use of \emph{any correct} semi-relativistic or relativistic method should bring the same result. The energy spectrum can be accurately established in any representation even if a resulting exponential transformation operator is not odd.

We can conclude that \emph{all} semi-relativistic and relativistic methods satisfying the Eriksen  conditions and based on an expansion of a transformation operator in a series asymptotically result to \emph{equivalent} FW Hamiltonians within regions of their applicability. This conclusion has been validated in Refs. \cite{erik,dVFor,PRA2016,ChiouChen}. In particular, the FW Hamiltonians for a spin-1/2 particle in electromagnetic fields obtained in Refs. \cite{JMP,RPJ} and Ref. \cite{ChiouChen} coincide despite a substantial difference in the methods of the FW transformation.  Our results presented below also support this conclusion.

The Eriksen conditions are satisfied automatically when one expands the Eriksen operator $U_{E}$ in a series of relativistic corrections
on powers of ${\cal O}/m$ and ${\cal E}/m$. Equations (\ref{eqXXI}) and (\ref{JMP2009}) can \emph{directly}
express the FW Hamiltonian as a series of such corrections
and are convenient for this purpose. This method applied in Ref. \cite{VJ} is semi-relativistic. One can use the formula
\begin{equation}\sqrt{{\cal H}^2}=
m\sqrt{1+\frac{{\cal H}^2-m^2}{m^2}}=
m\sqrt{1+\frac{2\beta m{\cal E}+{\cal O}^2+{\cal E}^2+
\{{\cal O},{\cal E}\}}{m^2}} \label{forkorn}
\end{equation}
and expand $({\cal H}^2)^{-1/2}$ in a Taylor series \cite{E,VJ}. While
needed calculations are cumbersome, they can be made analytically
with a computer \cite{VJ}. It has been supposed that ${\cal O}/(mc^2)\sim(v/c)$ and ${\cal
E}/(mc^2)\sim(v/c)^2$. We present below the exact FW Hamiltonian
calculated by de Vries and Jonker \cite{VJ} up to terms of the order of
$(v/c)^8$ ($m\alpha^8$ for light atoms). The result of calculations can be
presented in a more convenient form \cite{TMPFW} via multiple
commutators. While the Eriksen operator (\ref{eqXXI}) is inapplicable in the
nonstationary case (it does not diagonalize the wave function), one can use the property resulting from Eq. (\ref{taeq3}) and write the FW Hamiltonian in the form
\begin{equation}\begin{array}{c}
{\cal H}_{FW}=\beta\left(m+ \frac {{\cal
O}^2}{2m}-\frac{{\cal O}^4}{8m^3}+\frac{{\cal
O}^6}{16m^5}-\frac{5{\cal O}^8}{128m^7}\right)\\+{\cal
E}-\frac{1}{128m^6}\left\{(8m^4-6m^2{\cal O}^2+5{\cal O}^4),[{\cal
O},[{\cal O},{\cal
F}]]\right\}\\+\frac{1}{512m^6}\left\{(2m^2-{\cal O}^2),[{\cal
O}^2,[{\cal O}^2,{\cal F}]]\right\}\\
+\frac{1}{16m^3}\beta\left\{{\cal O},\left[[{\cal O},{\cal
F}],{\cal F}\right]\right\}-\frac{1}{32m^4}\left[{\cal
O},\left[[[{\cal O},{\cal F}],{\cal F}],{\cal
F}\right]\right]\\+\frac{11}{1024m^6}\left[{\cal O}^2,\left[{\cal
O}^2,[{\cal O},[{\cal O},{\cal F}]]\right]\right]+A_{24},
\end{array} \label{eq12erk}
\end{equation}
where 
\begin{equation}\begin{array}{c} A_{24}=\frac{1}{256m^5}\beta\Biggl(24\left\{{\cal O}^2, ([{\cal O},{\cal F}])^2\right\}-
20\left ([{\cal O}^2,{\cal F}]\right )^2-14\left\{{\cal O}^2,
\left [[{\cal O}^2,{\cal F}],{\cal F}\right]
\right\}\\-4\left[{\cal O},\left[{\cal O},\left[[{\cal O}^2,{\cal
F}],{\cal F}\right] \right] \right]+\frac92 \left[\left[{\cal
O},\left[{\cal O},\left[{\cal O}^2,{\cal F}\right]\right]
\right],{\cal F}\right]\\-\frac92 \left[\left[{\cal O},\left[{\cal
O},{\cal F}\right]\right],\left[{\cal O}^2,{\cal F}\right]\right]
+\frac52 \left [{\cal O}^2,\left[{\cal O},\left[[{\cal O},{\cal
F}],{\cal F}\right] \right] \right]\Biggr).
\end{array} \label{eq12A}
\end{equation}
In $A_{24}$, the first and second subscripts indicate the
respective numbers of ${\cal F}$ and ${\cal O}$ operators in the
product. A mistake in the calculation of this term made in Ref.
\cite{TMPFW} has been corrected in Ref. \cite{PRA2015}. In Ref. \cite{PRA2016}, the semi-relativistic FW Hamiltonian has been presented in the nonstationary case. In relation to light atoms, Eqs. (\ref{eq12erk}) and
(\ref{eq12A}) contain all terms up to the order of $m\alpha^8$ originating from the initial Dirac or Dirac-Pauli equation. We should note that the Hamiltonian may be written down differently in terms of commutators and anticommutators. Regrouping the terms in the Hamiltonian may lead to different sets of commutators and anticommutators.

Evidently, Eqs. (\ref{eq12erk}) and (\ref{eq12A}) as well as subsequent equations cover the nonstationary case. Their generalization is made on the base of Eq. (\ref{taeq3}). We suppose that nevertheless the Eriksen method can be rigorously generalized to the nonstationary case and consider this task as an outlook. In the stationary case, the Eriksen method is the simplest and most reliable semi-relativistic method of the FW transformation. This method directly leads to the general and easy-to-use equation for the FW Hamiltonian. Not only terms utilized in Eqs. (\ref{eq12erk}) and (\ref{eq12A}) but also terms of higher orders up to $(v/c)^{12}$ have been calculated many years ago (see Ref. \cite{VJ} and references therein). 

The Eriksen method finalized by de Vries and Jonker \cite{E,VJ} allows one to derive the FW Hamiltonian \emph{directly} via a simple substitution of operators ${\cal F}$ and ${\cal O}$ in the final formulas like Eqs. (\ref{eq12erk}) and (\ref{eq12A}) while other correct semi-relativistic methods need a precedent determination of the odd resulting exponential operator and cumbersome subsequent derivations. Formulas more precise formulas than Eqs. (\ref{eq12erk}) and (\ref{eq12A}) can be obtained using a computer.

Semi-relativistic methods of the FW transformation are actively used in the framework of nonrelativistic quantum electrodynamics (NRQED); see Refs. \cite{Pachucki,Pachucki2006,JentschuraCzarneckiPachucki,RHill,PatkosPachucki,Mei2014,Qiao2019,Zhou2023,ZhouJPhysB2023,Jentschura2024} and references therein. NRQED has been developed by Caswell and Lepage \cite{CaswLep} with the use of scattering matching. We add to NRQED the effective-field theory which has been elaborated by Pachucki \emph{et al.} \cite{Pachucki,Pachucki2006,JentschuraCzarneckiPachucki} and applies the FW transformation. Contrary to \emph{relativistic} quantum chemistry, NRQED considers light atoms. For such atoms, a high precision can be reached by using semi-relativistic methods of the FW transformation in conjunction with QED corrections obtained in the nonrelativistic approximation. In NRQED, different transformation methods are used and obtained FW Hamiltonians also differ. An important analysis has been fulfilled in Refs. \cite{Zhou2023,Jentschura2024}. It has been stated in Ref. \cite{Jentschura2024} that there are three different expressions for the terms in the FW Hamiltonian which are proportional to $\alpha^6$ $(\alpha=e^2/(\hbar c)\approx1/137)$. Our precedent analysis clearly shows that this is because some transformation methods do not satisfy the Eriksen conditions (\ref{VveEfrt}) and (\ref{Erikcon}). For this reason, applying the iterative method by Foldy and Wouthuysen \cite{FW} without its correction does not lead to the FW representation. Unfortunately, this method has been used in many papers and its implementation has even been recommended in Ref. \cite{Jentschura2024}. Contrary to such papers, Refs. \cite{Pachucki,Pachucki2006,JentschuraCzarneckiPachucki,Mei2014,Zhou2023} are based on the resulting exponential transformation operator  
\begin{equation}\begin{array}{c} {\cal S}_{res}=-\frac{i}{2m}\left\{\beta{\cal O}-\frac{1}{3m^2}\beta{\cal O}^3+\frac{1}{2m}\left[{\cal O},{\cal E}-i\hbar\frac{\partial}{\partial t}\right]+Y\right\}\\=-\frac{i}{2m}\left\{\beta\bm\alpha\cdot\bm\pi-\frac{1}{3m^2}\beta(\bm\alpha\cdot\bm\pi)^3+\frac{1}{2m}\left[\bm\alpha\cdot\bm\pi,eA_0-i\hbar\frac{\partial}{\partial t}\right]+Y\right\},
\end{array} \label{eqM}
\end{equation} where $\bm\alpha$ is a Dirac matrix, $\bm\pi=\bm p-e\bm A$ is the operator of the kinetic momentum and $Y$ is an unspecified odd operator. The operator $Y$ should cancel all high-order odd terms at the end of the calculation. Thus, only
one FW transformation with the operator (\ref{eqM}) is required to obtain the FW Hamiltonian and the operator ${\cal S}_{res}$ is odd and Hermitian. As a result, the transformation method used in Refs. \cite{Pachucki,Pachucki2006,JentschuraCzarneckiPachucki,Mei2014,Zhou2023} is quite correct and should lead to the FW representation. Certainly, one can also take into consideration terms in the Dirac-Pauli Hamiltonian (see Eq. (\ref{eqDiP}) below) describing the anomalous magnetic moment of the electron (see Refs. \cite{JentschuraCzarneckiPachucki,Zhou2023} and Eq. (\ref{equatEO}) below). Significant discrepancies in final Hamiltonians obtained with this method and the iterative one by Foldy and Wouthuysen \cite{FW} have been specified in Ref. \cite{Zhou2023}. In this connection, we underline that \emph{there is no ambiguity} in the FW Hamiltonians because the correct FW transformation defined by the Eriksen conditions (\ref{VveEfrt}) and (\ref{Erikcon}) is \emph{unique}. The use of false FW Hamiltonians brings inaccuracies in magnetic moments of atoms and other atomic parameters.

In Ref. \cite{ZhouJPhysB2023}, the \emph{relativistic} FW transformation has been fulfilled by the Douglas-Kroll-Hess approach. The use of this relativistic approach in NRQED is not necessary but can be helpful.

The Eriksen method was not practically applied for relativistic particles because the exact equations (\ref{eq3X3}) and (\ref{eqXXI}) contain the square roots
of Dirac matrices. In the present study, this method is used in the relativistic case for the first time. Formerly, a possibility to derive a relativistic FW Hamiltonian have been realized with other relativistic methods. There are a lot of such methods (see Refs. \cite{PRA2016,ChiouChen} and references therein). But only the method developed in Refs. \cite{PRA2015,PRA,JMP} allows one to obtain the \emph{relativistic} FW Hamiltonian, in which terms of the zeroth and first orders in the Planck constant and such terms of the order of $\hbar^2$ which describe contact interactions are \emph{exact}. Other terms are not specified. The first iteration has been carried out \cite{PRA} with the operator 
\begin{equation}
U=\frac{\epsilon+{\cal M}+\beta{\cal
O}}{\sqrt{(\beta\epsilon+\beta {\cal M}-{\cal O})^2}}.
\label{eq18N} \end{equation} For Dirac particles, ${\cal M}$ should be replaced with $mc^2$.
The corresponding Hamiltonian is valid in arbitrarily strong external fields and is given by \cite{PRA2015,PRA} 
\begin{equation}
\begin{array}{c}
{\cal H}_{FW}=\beta\epsilon'+ {\cal E}+\frac 14\left\{\frac{1}
{2{\epsilon'}^2+\{{\epsilon'},{\cal M}\}},\left(\beta\left[{\cal O},[{\cal O},{\cal
M}]\right]-[{\cal O},[{\cal O},{\cal
F}]]\right)\right\}, \quad \epsilon'=\sqrt{{\cal M}^2+{\cal
O}^2}.
\end{array}
\label{MHamf}
\end{equation}

We can check an agreement between Eqs. (\ref{eq12erk}) and (\ref{MHamf}). For Dirac particles, ${\cal M}=mc^2,~[{\cal M},{\cal O}]=0$, and Eq.  (\ref{MHamf}) takes the form \cite{JMP,PRA2015}
\begin{equation}
\begin{array}{c}
{\cal H}_{FW}=\beta\epsilon+ {\cal E}-\frac 14\left\{\frac{1}
{2{\epsilon}({\epsilon}+m)},[{\cal O},[{\cal O},{\cal
F}]]\right\}, \quad \epsilon=\sqrt{{m}^2+{\cal
O}^2}
\end{array}
\label{MHamD}
\end{equation}
and the series expansion of Eq. (\ref{MHamD}) is given by
\begin{equation}\begin{array}{c}
{\cal H}_{FW}=\beta\left(m+ \frac {{\cal
O}^2}{2m}-\frac{{\cal O}^4}{8m^3}+\frac{{\cal
O}^6}{16m^5}-\frac{5{\cal O}^8}{128m^7}\right)\\+{\cal
E}-\frac{1}{128m^6}\left\{(8m^4-6m^2{\cal O}^2+5{\cal O}^4),[{\cal
O},[{\cal O},{\cal
F}]]\right\}.
\end{array} \label{eq12}
\end{equation}

Evidently, Eqs. (\ref{eq12erk}) and (\ref{eq12}) agree. However, the latter equation does not cover all terms linear in ${\cal F}$. A determination of such terms is 
presented in next sections. For such a determination, it is difficult to use the relativistic approach developed in Refs. \cite{JMP,PRA2015,PRA} and other publications
(see Refs. \cite{PRA2016,ChiouChen} and references therein). Taking into account additional terms needs successive iterations. However, an implementation of such iterations requires serious efforts just in the relativistic case because necessary calculations become rather cumbersome. The reason for this is the necessity to satisfy the Eriksen condition (\ref{Erikcon}) on each step of the iterations. Therefore, in this paper we do not use iterative methods and apply the exact FW transformation operator (\ref{eqXXI}) for the derivation of a leading correction.

We also mention the so-called exact FW transformation \cite{erik,exactFW} which often allows one to obtain exact block-diagonalized Hamiltonians but does not lead to the FW representation.

\section{Operator extraction of a square root} \label{Extraction}

In the present study, we derive the relativistic FW Hamiltonian in the weak-field approximation. We fulfill needed calculations for the stationary case with the use of the Eriksen operator (\ref{eqXXI}) and then pass to the nonstationary case following Eq. (\ref{taeq3}).

The well-known FW transformation for a free Dirac particle clearly shows \cite{FW} that the commutators like $[{\cal O},[{\cal O},...[{\cal O},{\cal
E}]...]]$ and $[{\cal O},[{\cal O},...[{\cal O},{\cal
F}]...]]$ are proportional to $\hbar$ appearing after the first commutation with respect to coordinates or time ($[{\cal O},{\cal
E}],~[{\cal O},{\cal
F}]$). Other commutations may not add factor $\hbar$ due to a noncommutativity of related odd Dirac matrices. On the contrary, the operator ${\cal O}^2$ is even and the commutators $[{\cal O}^2,[{\cal O}^2,{\cal
E}]]$ and $[{\cal O}^2,[{\cal O}^2,{\cal
F}]]$ are proportional to $\hbar^2$. The commutators $[{\cal O}^2,[{\cal O}^2,[{\cal
O},[{\cal O},{\cal
E}]]]]$ and $[{\cal O}^2,[{\cal O}^2,[{\cal
O},[{\cal O},{\cal
F}]]]]$ are proportional to $\hbar^3$. In the present study, we determine only terms proportional to $\hbar^2$ and use the weak-field approximation.

Previous applications of the Eriksen operator (\ref{eqXXI}) utilized the expansion (\ref{forkorn}) of $({\cal H}^2)^{-1/2}$. We apply an operator extraction of a square root in the weak-field approximation. The particle momentum connected to ${\cal O}$ can be arbitrarily large. The obtained relation should give $({\cal H}^2)$ after squaring. Therefore, the result obtained should be nontrivial and should include some term or terms caused by a noncommutativity of operators. In the weak-field approximation, 
\begin{equation}\begin{array}{c}
{\cal H}^2=\epsilon^2 +\left\{\beta m+{\cal
O},{\cal
E}\right\},\qquad  \epsilon=\sqrt{m^2+{\cal O}^2},\\ \sqrt{{\cal H}^2}=\epsilon
+\frac{1}{4}\left\{\frac{1}{\epsilon},\left\{\beta m+{\cal
O},{\cal
E}\right\}\right\}-\frac{1}{8}\left\{\frac{\beta m+{\cal
O}}{\epsilon^3},[\epsilon,[\epsilon,{\cal
E}]]\right\}\\ 
\approx \epsilon
+\frac{1}{4}\left\{\frac{1}{\epsilon},\left\{\beta m+{\cal
O},{\cal
E}\right\}\right\}-\frac{1}{32}\left\{\frac{\beta m+{\cal
O}}{\epsilon^5},[{\cal O}^2,[{\cal O}^2,{\cal
E}]]\right\}.
\end{array} \label{eq7}
\end{equation}
The operators $\beta m+{\cal O}$ and $\epsilon$ commute. 
Therefore, 
\begin{equation}\begin{array}{c}
\frac{1}{\sqrt{{\cal H}^2}}=\frac1\epsilon
-\frac{1}{4\epsilon}\left\{\frac{1}{\epsilon},\left\{\beta m+{\cal
O},{\cal
E}\right\}\right\}\frac{1}{4\epsilon}+\frac{1}{8}\left\{\frac{\beta m+{\cal
O}}{\epsilon^5},[\epsilon,[\epsilon,{\cal
E}]]\right\}\\ 
\approx \frac1\epsilon
-\frac{1}{4\epsilon}\left\{\frac{1}{\epsilon},\left\{\beta m+{\cal
O},{\cal
E}\right\}\right\}\frac{1}{4\epsilon}+\frac{1}{32}\left\{\frac{\beta m+{\cal
O}}{\epsilon^7},[{\cal O}^2,[{\cal O}^2,{\cal
E}]]\right\}.
\end{array} \label{rever}
\end{equation}
Our approach can be used for taking into account terms quadratic (bilinear) in ${\cal E}$. This problem is solved with the use of the well-known formulas $$\sqrt{1+x}=1+\frac x2-\frac{x^2}{8}+\dots,\qquad \frac{1}{\sqrt{1+x}}=1-\frac x2+\frac{3x^2}{8}-\dots,$$ but series expansions should include terms appearing owing to the noncommutativity of the operators $\epsilon^2$ and $\left\{\beta m+{\cal O},{\cal E}\right\}$. As in Eqs. (\ref{eq7}) and (\ref{rever}), such terms should be determined by squaring.

The formulas obtained allow us to determine $\lambda$. It is convenient to calculate it in the form
$$\lambda=\frac12\left\{{\cal H},({\cal H}^2)^{-1/2}\right\}.$$ The calculation results in
\begin{equation}\begin{array}{c}
\lambda=\frac{\beta m+{\cal
O}}{\epsilon}-\frac{1}{8}\left\{\frac{1}{\epsilon^3},[\epsilon,[\epsilon,{\cal
E}]]\right\}+\frac{1}{8}\left\{\frac{1}{\epsilon^3},[\beta m+{\cal
O},[\beta m+{\cal
O},{\cal E}]]\right\}\\ \approx \frac{\beta m+{\cal
O}}{\epsilon}-\frac{1}{32}\left\{\frac{1}{\epsilon^5},[{\cal O}^2,[{\cal O}^2,{\cal
E}]]\right\}+\frac{1}{8}\left\{\frac{1}{\epsilon^3},[{\cal
O},[{\cal
O},{\cal E}]]\right\}+\frac{\beta m}{4}\left\{\frac{1}{\epsilon^3},[{\cal
O},{\cal E}]\right\}. 
\end{array} \label{lambd}
\end{equation}
It follows from Eq. (\ref{lambd}) that 
\begin{equation}\begin{array}{c}
2+\beta\lambda+\lambda\beta=\frac{2(\epsilon+m)}{\epsilon}-\frac{\beta}{16}\left\{\frac{1}{\epsilon^5},[{\cal O}^2,[{\cal O}^2,{\cal
E}]]\right\}+\frac{\beta}{4}\left\{\frac{1}{\epsilon^3},[{\cal
O},[{\cal 
O},{\cal E}]]\right\}.
\end{array} \label{sqH}
\end{equation} 

Analyzing Eqs. (\ref{eq12erk}) and (\ref{eq12A}) as well as the corresponding equations in Refs. \cite{Zhou2023,ZhouJPhysB2023,Jentschura2024}, we can conclude that a derivation of relativistic terms quadratic in ${\cal E}$ allows one to determine all nonzero terms up to the order of $m\alpha^8$ in NRQED. It follows from this section that the derivation of relativistic terms quadratic in ${\cal E}$ is possible. While this derivation can be rather cumbersome, it can be considered as an outlook and an objective of future studies.

\section{Relativistic Foldy-Wouthuysen Hamiltonian} \label{FWH}

The next derivation of the relativistic FW Hamiltonian is straightforward. In the stationary case,
\begin{equation}\begin{array}{c} {\cal H}_{FW}=U_E{\cal H}U_E^{-1}=\frac{1}{\sqrt{2+\beta\lambda+\lambda\beta}}\left(1+\beta\lambda\right){\cal H}\left(1+\lambda\beta\right)\frac{1}{\sqrt{2+\beta\lambda+\lambda\beta}}\\=\frac{1}{\sqrt{2+\beta\lambda+\lambda\beta}}\left[2(\beta m+{\cal E})+\beta\sqrt{{\cal H}^2}+\sqrt{{\cal H}^2}\beta\right]\frac{1}{\sqrt{2+\beta\lambda+\lambda\beta}}.
\end{array} \label{taueq}
\end{equation}

Evidently, this Hamiltonian is even.
It follows from Eq. (\ref{sqH}) that
\begin{equation}\begin{array}{c}
\frac{1}{\sqrt{2+\beta\lambda+\lambda\beta}}=\sqrt{\frac{\epsilon}{2(\epsilon+m)}}\\+\frac{\beta}{128}\left\{\sqrt{\frac{\epsilon}{2(\epsilon+m)}},\left(\left\{\frac{1}{\epsilon^4(\epsilon+m)},[{\cal O}^2,[{\cal O}^2,{\cal
E}]]\right\}-\left\{\frac{4}{\epsilon^2(\epsilon+m)},[{\cal
O},[{\cal
O},{\cal E}]]\right\}\right)\right\}.
\end{array} \label{sqrte}
\end{equation}

The calculations result in
\begin{equation}
\begin{array}{c}
{\cal H}_{FW}=\beta\epsilon+ {\cal E}-\frac 18\left\{\frac{1}
{\epsilon(\epsilon+m)},[{\cal O},[{\cal O},{\cal
E}]]\right\}+\frac{1}{64}\left\{\frac{2\epsilon^2-m^2}{\epsilon^4(\epsilon+m)^2},[{\cal O}^2,[{\cal O}^2,{\cal
E}]]\right\}.
\end{array}
\label{Hamwf}
\end{equation}
We can now use the property following from Eq. (\ref{taeq3}) and present Eq. (\ref{Hamwf}) in the form
\begin{equation}
\begin{array}{c}
{\cal H}_{FW}=\beta\epsilon+ {\cal E}-\frac 18\left\{\frac{1}
{\epsilon(\epsilon+m)},[{\cal O},[{\cal O},{\cal
F}]]\right\}+\frac{1}{64}\left\{\frac{2\epsilon^2-m^2}{\epsilon^4(\epsilon+m)^2},[{\cal O}^2,[{\cal O}^2,{\cal
F}]]\right\}.
\end{array}
\label{Hamwfin}
\end{equation}
This equation demonstrates our main result. The first three terms have the same form for weak and arbitrarily strong external fields, but they do not give an exact FW Hamiltonian. The last term derived in the present work is new and defines the leading relativistic correction in the weak-field approximation. This term is proportional to $\hbar^2$ and exhaustively describes relativistic corrections proportional to $\hbar^2$ and linear in ${\cal F}$.

We can prove the full agreements between Eqs. (\ref{eq12erk}) and (\ref{Hamwfin}). The expansion of the factor $(2\epsilon^2-m^2)/[\epsilon^4(\epsilon+m)^2]$ in a series of relativistic corrections has the form
\begin{equation}\frac{2\epsilon^2-m^2}{\epsilon^4(\epsilon+m)^2}\approx\frac{2m^2-{\cal O}^2}{8m^6}.\label{mprne}
\end{equation}
The agreement with Eq. (\ref{eq12erk}) following from Eq. (\ref{mprne}) is an important confirmation of our result. The only remaining term in Eq. (\ref{eq12erk}) which is linear with respect to ${\cal F}$ is proportional to $\hbar^3$ (see Sec. \ref{Extraction}).

\section{Relativistic wave equation of the second order} \label{SecondOrderE}

Our correction to the previously obtained FW Hamiltonian (\ref{MHamD}) is proportional to $\hbar^2$. Nevertheless, Eq. (\ref{eq12erk}) shows the presence 
of other terms proportional to $\hbar^2$. The kinetic ($T=\epsilon-m$) and potential ($V={\cal
E}$) energies of electrons in atoms are of the same order of magnitude due to the virial theorem. In particular, $\langle T\rangle=\langle\bm p^2\rangle/(2m)=-\langle V\rangle/2$ and $\langle T\rangle\sim \langle |V|\rangle\sim m\alpha^2$ for nonrelativistic electrons in light atoms. For any atom, the precision should be much better than of the order of $m\alpha^2$. Therefore, the weak-field approximation is not applicable in quantum chemistry and NRQED. In NRQED, the precision of the order of $m\alpha^8$ should be reached. In quantum chemistry (including relativistic quantum chemistry), any appropriate precision can be obtained using a computer. In these cases, terms containing ${\cal
E}$ is the second or even higher degrees should be taken into account. However, this approximation can often be useful in nonrelativistic and relativistic scattering theories. Even if the impact parameter $\mathfrak{b}$ is of the order of the Compton wavelength, the condition of the weak-field approximation $|V|\ll cp$ can be satisfied (see Sec. \ref{Discussion}).

We should note that the FW representation is fully equivalent to the Schr\"{o}dinger one and extends the latter representation to the field of relativistic energies \cite{PRAFW}. Therefore, the application of the FW representation in the relativistic scattering theory is rather promising. However, the presence of the operator $\bm p^2$ under the square root (${\cal O}^2=\bm p^2$ for a free particle) causes some inconvenience. To overcome it, one can move to a second-order equation. The rigorous approach elaborated in Ref. \cite{PAN} can be used. It should be mentioned that simple squaring is inappropriate because it leads to an appearance of non-Hermitian terms (see Ref. \cite{PAN}). One needs to start from the second-order Klein-Gordon-like equation and to fulfill the subsequent generalized Feshbach-Villars transformation \cite{TMP2008}. To explain the method used, we report the results obtained in Ref. \cite{TMP2008}. We can write the second-order equation in the form \cite{TMP2008} \begin{equation}
\begin{array}{c}
\left[\left(i\frac{\partial}{\partial t}-{\cal A}_0\right)^2-(\bm p-\bm{\mathcal{A}})^2-m^2\right]\psi=0.
\end{array}
\label{eqGFV}
\end{equation} For spinning particles, the quantities ${\cal A}_0$ and $\bm{\mathcal{A}}$ can be spin-dependent. The generalized Feshbach-Villars transformation is based on the introduction of two wave functions, 
$\phi$ and $\chi$, which are defined by
\begin{equation}
\begin{array}{c}
\psi=\phi+\chi, \qquad \left(i\frac{\partial}{\partial t}-{\cal A}_0\right)\psi=N(\phi-\chi),
\end{array}
\label{eqnwf}
\end{equation} where $N$ is an arbitrary nonzero real parameter. It is used instead of the mass $m$ in the original Feshbach-Villars transformation \cite{FV}. The introduced wave functions form the bispinor-like wave function $\Psi=\left(\begin{array}{c}\phi \\ \chi\end{array}\right)$. If we multiply the last relation by $i\frac{\partial}{\partial t}-{\cal A}_0$, then we
can represent Eq. (\ref{eqnwf}) in the matrix form \cite{TMP2008}:
\begin{equation}
\begin{array}{c}
i\frac{\partial\Psi}{\partial t}={\cal H}\Psi, \qquad {\cal H}=\rho_3\frac{\bm\pi^2+m^2+N^2}{2N}+ {\cal A}_0+i\rho_2\frac{\bm\pi^2+m^2-N^2}{2N},
\end{array}
\label{eqnmf}
\end{equation} where $\bm\pi=\bm p-\bm{\mathcal{A}}=-i\nabla-\bm{\mathcal{A}}$ is the kinetic momentum operator and $\rho_i~(i=1,2,3)$ are the Pauli matrices, whose
components act on the corresponding components of the wave function $\Psi$. 

In the Hamiltonian (\ref{eq3G}), we obtain
\begin{equation} {\cal M}=\frac{\bm\pi^2+m^2+N^2}{2N},\qquad{\cal E}={\cal A}_0,
\qquad {\cal O}=i\rho_2\frac{\bm\pi^2+m^2-N^2}{2N}. \label{eqpef} \end{equation}
In this case, $\epsilon'=\sqrt{{\cal M}^2+{\cal O}^2}=\sqrt{m^2+\bm\pi^2}$. Since corrections are proportional to $\hbar^2$, we cannot use Eq. (\ref{MHamf}) which does not specify these corrections and should fulfill the FW transformation with the operator \cite{JMP,PRA,PRA2015,TMP2008}
\begin{equation}
\begin{array}{c}
U_{FW}=\frac{\epsilon'+{\cal M}+\beta{\cal O}}{\sqrt{2\epsilon'(\epsilon'+{\cal M})}},\qquad
U_{FW}^{-1}=\frac{\epsilon'+{\cal M}-\beta{\cal O}}{\sqrt{2\epsilon'(\epsilon'+{\cal M})}}.
\end{array}
\label{TMP2008o}
\end{equation}

To apply this method for the derivation of a second-order equation for spin-1/2 particles, we need to present Eq. (\ref{eqGFV}) in the form
\begin{equation}
\begin{array}{c}
\left[\left(i\frac{\partial}{\partial t}-\mathfrak{E}\right)^2-\mathfrak{O}^2-m^2\right]\psi=0.
\end{array}
\label{eqgenrl}
\end{equation}
Next, we obtain the equation for the Hamiltonian:
\begin{equation}
\begin{array}{c}
i\frac{\partial\Psi}{\partial t}={\cal H}\Psi, \qquad {\cal H}=\rho_3\frac{\mathfrak{O}^2+m^2+N^2}{2N}+\mathfrak{E}+i\rho_2\frac{\mathfrak{O}^2+m^2-N^2}{2N}.
\end{array}
\label{eqnml}
\end{equation}
In this case, the specification of terms in the Hamiltonian (\ref{eq3G}) and their substitution into Eq. (\ref{TMP2008o}) results in \cite{TMP2008}
\begin{equation} U_{FW}=\frac{\epsilon'+N+\rho_1(\epsilon'-N)}{2\sqrt{\epsilon'N}},\qquad
U_{FW}^{-1}=\frac{\epsilon'+N-\rho_1(\epsilon'-N)}{2\sqrt{\epsilon'N}},\qquad\epsilon'=\sqrt{m^2+\mathfrak{O}^2}.
\label{eqpenew} \end{equation}

After the transformation, the resulting \emph{approximate} FW Hamiltonian has the form \cite{TMP2008}
\begin{equation}
\begin{array}{c} {\cal H}_{FW}=\rho_3\epsilon'+{\cal E}'+{\cal O}',\qquad {\cal E}'=\mathfrak{E}+\frac{1}{2\sqrt{\epsilon'}}
\Bigl[\sqrt{\epsilon'},\bigl[\sqrt{\epsilon'},\mathfrak{F}\bigr]\Bigr]\frac{1}{\sqrt{\epsilon'}},\\ {\cal O}'=\frac{\rho_1}{2\sqrt{\epsilon'}}\bigl[\sqrt{\epsilon'},\mathfrak{F}\bigr]\frac{1}{\sqrt{\epsilon'}},\qquad
\mathfrak{F}=\mathfrak{E}-i\hbar\frac{\partial}{\partial t}.
\end{array}
\label{eqpenFW} \end{equation} Importantly, the operators ${\cal E}'$ and ${\cal O}'$ do not depend on $N$ \cite{TMP2008}.
The operator ${\cal O}'$ introduces the contribution proportional to ${\cal O}'^2$ into the final (even) FW Hamiltonian (see Ref. \cite{PRA2016}). Therefore, it can be disregarded. Approximately (see general formulas in Ref. \cite{JMP}),
$$
\begin{array}{c} 
\bigl[\sqrt{\epsilon'},\mathfrak{F}\bigr]=\frac14\left\{\frac{1}{\sqrt{\epsilon'}},\bigl[\epsilon',\mathfrak{F}\bigr]\right\}=\frac{1}{8}\left\{\frac{1}{\epsilon'\sqrt{\epsilon'}},\bigl[\mathfrak{O}^2,\mathfrak{F}\bigr]\right\},\\
\frac{1}{2\sqrt{\epsilon'}}
\Bigl[\sqrt{\epsilon'},\bigl[\sqrt{\epsilon'},\mathfrak{F}\bigr]\Bigr]\frac{1}{\sqrt{\epsilon'}}=\frac{1}{64}\left\{\frac{1}{{\epsilon'}^4},\Bigl[\mathfrak{O}^2,\bigl[\mathfrak{O}^2,\mathfrak{F}\bigr]\Bigr]\right\}
\end{array} $$
and
\begin{equation}
{\cal E}'=\mathfrak{E}+\frac{1}{64}\left\{\frac{1}{{\epsilon'}^4},\Bigl[\mathfrak{O}^2,\bigl[\mathfrak{O}^2,\mathfrak{F}\bigr]\Bigr]\right\}.
\label{eqpnlFW} \end{equation}

Our derivation does not use a specific form of the FW Hamiltonian and remains valid for particles with any spins.

Therefore, the quantity ${\cal E}'$ in the FW Hamiltonian defined by Eqs. (\ref{eqpenFW}) and (\ref{eqpnlFW}) corresponds to $\mathfrak{E}$ in the second-order equation (\ref{eqgenrl}). This connection between the two quantities is valid for particles with any spins. For Dirac particles, we find that ${\cal E}'$ is equal to ${\cal H}_{FW}-\beta\epsilon$ in the FW Hamiltonian (\ref{Hamwfin}). A comparison of other quantities in Eqs. (\ref{Hamwfin}) and (\ref{eqpnlFW}) results in $\mathfrak{O}={\cal O}$ and $\epsilon'=\epsilon$. We can now conclude that $\mathfrak{E}$ is approximately given by
\begin{equation} \begin{array}{c}
\mathfrak{E}={\cal E}-\frac 18\left\{\frac{1}
{\epsilon(\epsilon+m)},[{\cal O},[{\cal O},{\cal
F}]]\right\}+\frac{1}{64}\left\{\frac{2\epsilon^2-m^2}{\epsilon^4(\epsilon+m)^2},[{\cal O}^2,[{\cal O}^2,{\cal
F}]]\right\}\\-\frac{1}{64}\left\{\frac{1}{\epsilon^4},[{\cal O}^2,[{\cal O}^2,{\cal
F}]]\right\}={\cal E}-\frac 18\left\{\frac{1}
{\epsilon(\epsilon+m)},[{\cal O},[{\cal O},{\cal
F}]]\right\}\\+\frac{1}{64}\left\{\frac{\epsilon^2-2(\epsilon+m)m}{\epsilon^4(\epsilon+m)^2},[{\cal O}^2,[{\cal O}^2,{\cal
F}]]\right\}.
\end{array} \label{eqpepnl} \end{equation}

The final form of the second-order equation for Dirac particles reads
\begin{equation}
\begin{array}{c}
\left[\left(i\frac{\partial}{\partial t}-\mathfrak{E}\right)^2-{\cal O}^2-m^2\right]\psi=0,
\end{array}
\label{eqgenrf}
\end{equation}
where $\mathfrak{E}$ is defined by Eq. (\ref{eqpepnl}).

\section{Examples: Dirac particle in electric and magnetic fields} \label{Example}

\subsection{Relativistic wave equations of the first and second orders for a Dirac particle in an electrostatic field} \label{Examplo}

The consideration of a Dirac particle (without an anomalous magnetic moment) in an electrostatic field gives one a practically important example of an application of derived formulas. In this case, the Dirac Hamiltonian reads (${\cal E}=e\Phi,~{\cal O}=\bm\alpha\cdot\bm p$)
\begin{equation} {\cal H}=\beta m+\bm\alpha\cdot\bm p+e\Phi,
\label{eqDirac} \end{equation} 
where $e$ is the positive or negative charge of the particle and $\Phi$ is the scalar potential of the electrostatic field. As follows from Eq. (\ref{Hamwfin}), the FW Hamiltonian has the form
\begin{equation}\begin{array}{c} {\cal H}_{FW}=\beta\epsilon+e\Phi+\frac
   {\mu_0m}{4}\left\{\frac{1}{\epsilon(\epsilon
   +m)},\Bigl[\bm\Sigma\cdot(\bm p
\times\bm E-\bm E\times\bm p)-\hbar\nabla
\cdot\bm E\Bigr]\right\}\\+\frac{e\hbar^2}{16}\left\{\frac{2\epsilon^2-m^2}{\epsilon^4(\epsilon+m)^2},(\bm{p}\cdot\nabla)(\bm{p}\cdot\bm E)\right\},
\end{array} \label{eq33new} \end{equation}
where $\mu_0=e\hbar/(2m)$ is the Dirac magnetic moment, $\bm E=-\nabla\Phi$ is the electric field, and $\epsilon=\sqrt{m^2+\bm{p}^2}$. It has been taken into account that $$[{\cal O}^2,[{\cal O}^2,{\cal
F}]]=-4e\hbar^2(\bm{p}\cdot\nabla)(\bm{p}\cdot\nabla)\Phi=4e\hbar^2(\bm{p}\cdot\nabla)(\bm{p}\cdot\bm E).$$

The results presented in this subsection are applicable to any electrostatic field (from a
point charge, uniform, and so on). However, the scattering theory does not consider the trivial case of a uniform electric field. Scattering of electrons (positrons, muons) by nuclei is the most important research topic. Nuclei can create either exactly a Coulomb field (helion) or a Coulomb field with a quadrupole component (deuteron). The field strength is limited by the weak-field approximation ($|e\Phi|\ll mc^2$).

The first-order equation (\ref{eq33new}) allows one to derive the corresponding second-order equation which is given by
\begin{equation}
\begin{array}{c}
\left[\left(i\frac{\partial}{\partial t}-\mathfrak{E}\right)^2-\bm{p}^2-m^2\right]\psi=0,\\
\mathfrak{E}=e\Phi+\frac
   {\mu_0m}{4}\left\{\frac{1}{\epsilon(\epsilon
   +m)},\Bigl[\bm\Sigma\cdot(\bm p
\times\bm E-\bm E\times\bm p)-\hbar\nabla
\cdot\bm E\Bigr]\right\}\\+\frac{e\hbar^2}{16}\left\{\frac{\epsilon^2-2(\epsilon+m)m}{\epsilon^4(\epsilon+m)^2},(\bm{p}\cdot\nabla)(\bm{p}\cdot\bm E)\right\}.
\end{array}
\label{eqgenef}
\end{equation}
A relative importance of corrections introduced by the last terms in Eqs. (\ref{eq33new}) and (\ref{eqgenef}) is defined by their comparison with the spin-dependent terms and is characterized by the ratio $\hbar/(rp)$, where $r\sim |\partial E_i/\partial x_j|/E$.

\subsection{Spin-1/2 particle with an anomalous magnetic moment in nonstationary electric and magnetic fields} \label{Examplt}

It is instructive to consider the contribution of the determined correction to the relativistic FW Hamiltonian for a spin-1/2 (Dirac-Pauli \cite{Pauli}) particle with an anomalous magnetic moment interacting with nonstationary electric and magnetic fields. The initial Hamiltonian is given by 
\begin{equation} {\cal H}=\beta m+\bm\alpha\cdot\bm\pi+e\Phi+\mu'(-\bm\Pi\cdot\bm B+i\bm\gamma\cdot\bm E),
\label{eqDiP} \end{equation}
where $\mu'$ is the anomalous magnetic moment and $\bm\Pi$ and $\bm\gamma$ are Dirac matrices. In this case
\begin{equation} {\cal E}=e\Phi-\mu'\bm\Pi\cdot\bm B,\qquad {\cal O}=\bm\alpha\cdot\bm\pi+i\mu'\bm\gamma\cdot\bm E.\label{equatEO} \end{equation}

In the weak-field approximation, the FW Hamiltonian previously obtained without the correction defined by the last term in Eq. (\ref{Hamwfin}) reads \cite{JMP}
\begin{equation}\begin{array}{c} 
{\cal H}_1=\beta\epsilon'+e\Phi+\frac
   14\left\{\left(\frac{\mu_0m}{\epsilon'
   +m}+\mu'\right)\frac{1}{\epsilon'},\Bigl[\bm\Sigma\cdot(\bm\pi
\times\bm E-\bm E\times\bm\pi)-\hbar\nabla
\cdot\bm E\Bigr]\right\}\\ -\frac
12\left\{\left(\frac{\mu_0m}{\epsilon'}
+\mu'\right), \bm\Pi\!\cdot\!\bm B\right\}\\
+\beta\frac{\mu'}{4}\left\{\frac{1}{\epsilon'(\epsilon'+m)},
\Bigl[(\bm{B}\!\cdot\!\bm\pi)(\bm{\Sigma}\!\cdot\!\bm\pi)+ (\bm{\Sigma}
\!\cdot\!\bm\pi)(\bm\pi\!\cdot\!\bm{B})+2\pi\hbar(\bm\pi\!\cdot\!\bm j+
\bm j\!\cdot\! \bm\pi)\Bigr]\right\},
\end{array} \label{eqn3new} \end{equation}
where $\epsilon'=\sqrt{m^2+\bm{\pi}^2}$ and
$$\bm j=\frac{1}{4\pi}\left(c\,\nabla\!\times\!\bm B-\frac{\partial \bm E}
{\partial t}\right)$$
is the density of the external electric current. This Hamiltonian covers the case of \emph{nonstationary} electromagnetic fields \cite{JMP}.

The relativistic Darwin interaction is presented by the term
proportional to $\nabla\cdot\bm E$. It is contributed by both
the Dirac and anomalous magnetic moments. 
The term proportional to $\bm j$ describes a similar
contact interaction with external electric currents.

The expression for the Darwin term in the FW representation is similar to the classical
formula defining the energy of a static interaction of an extended
particle with external charges: $W=-(e/6)\langle r_0^2\rangle\nabla\cdot\bm E$.
So, the Darwin interaction defines the \emph{effective} root-mean-square
radius of a point-like particle, $\langle r_0^2\rangle$.

In the approximation used, 
\begin{equation}\begin{array}{c} {\cal O}^2= m^2+\bm{\pi}^2-e\bm\Sigma\cdot\bm B+\beta\mu'
\left[\bm\Sigma\cdot(\bm\pi\times\bm E)-\bm\Sigma\cdot(\bm
E\!\times \bm\pi)-\nabla\cdot\bm E\right],\\
\left[{\cal O}^2,{\cal F}\right]=ie(\bm\pi\cdot\bm E+\bm E\cdot\bm\pi)-ie\bm\Sigma\cdot\dot{\bm B}+i\mu'\left[\bm\pi\cdot\bigl((\bm\Pi\cdot\nabla)\bm B\bigr)+\bigl((\bm\Pi\cdot\nabla)\bm B\bigr)\cdot
\bm\pi\right]\\-4\pi i\mu'[\bm\Pi\cdot(\bm\pi\times\bm j)-\bm\Pi\cdot(\bm j\times\bm\pi)],
\end{array} \label{eqOse} \end{equation}
where the dot indicates the time derivative. For the derivation, we have utilized well-known equations for the electromagnetic fields and potentials. The term proportional to $\nabla\cdot\bm E$ has been neglected. Importantly, derivatives of the potentials enter Eq. (\ref{eqOse}) only implicitly (via the fields $\bm E$ and $\bm B$).
The newly obtained correction is equal to
\begin{equation}
\begin{array}{c}
{\cal H}_{2}=\frac{1}{64}\left\{\frac{2{\epsilon'}^2-m^2}{{\epsilon'}^4({\epsilon'}+m)^2},{\cal K}\right\},\qquad {\cal K}=[{\cal O}^2,[{\cal O}^2,{\cal
F}]]=-i(\bm\pi\cdot\nabla)[{\cal O}^2,{\cal F}]-i(\nabla[{\cal O}^2,{\cal F}])\cdot\bm\pi.
\end{array}
\label{Hamwatn}
\end{equation}
The resulting relativistic nonstationary FW Hamiltonian is equal to
\begin{equation}
{\cal H}_{FW}={\cal H}_{1}+{\cal H}_{2}.
\label{HamFW}
\end{equation}

We can evaluate the correction defined by Eqs. (\ref{eqOse}) and (\ref{Hamwatn}). This correction should be compared with the main spin-independent and spin-dependent terms in the total FW Hamiltonian. The comparison shows that such terms are defined by the operator ${\cal E}$. The correction is nonzero only for nonuniform electric and magnetic fields. In particular, such fields can be created by an electric quadrupole with the scalar potential $\Phi=V(2z^2-\rho^2)$ and the magnetic bottle with the field $$\bm B=B_2\left[\left(z^2-\frac{\rho^2}{2}\right)\bm e_z-z\rho\bm e_\rho\right].$$
Here $V=const,\,B_2=const$. 

Let $d$ be the size characterizing the area influenced by a nonuniform field. In this case,
$$\begin{array}{c}
\frac14\Bigl\{(\bm\pi\cdot\nabla)(\bm\pi\cdot\bm E+\bm E\cdot\bm\pi)+\nabla(\bm\pi\cdot\bm E+\bm E\cdot\bm\pi)\cdot\bm\pi \Bigr\}\sim\frac{|\bm\pi|^2E}{d}\sim\frac{|\bm\pi|^2\Phi}{d^2},\\
\frac14\Bigl\{(\bm\pi\cdot\nabla)\left[\bm\pi\cdot\bigl((\bm\Pi\cdot\nabla)\bm B\bigr)+\bigl((\bm\Pi\cdot\nabla)\bm B\bigr)\cdot
\bm\pi\right]+\nabla\bigl[\bm\pi\cdot\bigl((\bm\Pi\cdot\nabla)\bm B\bigr)\\ +\bigl((\bm\Pi\cdot\nabla)\bm B\bigr)\cdot
\bm\pi\bigr]\cdot\bm\pi\Bigr\}
\sim\frac{|\bm\pi|^2B}{d^2}.
\end{array}$$
Evidently, the correction ${\cal H}_{2}$ is not small compared with ${\cal H}_{1}$ when a particle is relativistic and $d|\bm\pi|\sim\hbar$. This situation can take place at scattering. 

However, the quantity ${\cal H}_{2}$ can usually be neglected even in high-precision experiments when electric and magnetic fields are created by macroscopic devices. For a relativistic particle, $$|{\cal H}_{2}|\sim\frac{1}{{\epsilon'}^2d^2}|{\cal H}_{1}|=\frac{\lambdabar^2}{\gamma^2d^2}|{\cal H}_{1}|,$$ where $\gamma$ is the Lorentz factor and $\lambdabar=\hbar/(mc)$ is the Compton wavelength. For the electron and positron, $\lambdabar=3.86\times 10^{-13}$ m. In the electron $g-2$ experiment fulfilled in the Penning trap, the electric field with $\Phi_0=5$ V and $V=1.56\times10^5$ V/m$^2$ has been used \cite{Dehmelt}. In this case, $d^2\sim\Phi_0/V\sim10^{-5}$ m$^2$. The relative error conditioned by neglecting ${\cal H}_{2}$ is of the order of $\lambdabar^2/d^2\sim10^{-20}$ or smaller and can be disregarded. For the magnetic bottle used in the Penning trap in Ref. \cite{Mooser}, $B=5.3$ T, $B_2=2.97\times10^5$ T/m$^2$. The corresponding relative error for the electron is also of the order of $\lambdabar^2/d^2$ or smaller. In this case, $d^2\sim B/B_2\sim10^{-5}$ m$^2$. Therefore, the corresponding relative error is of the same order of magnitude (no more than $10^{-20}$) and can also be neglected.

The considered examples show when the correction obtained in Sec. \ref{FWH} and considered in this section can be important. The correction is valid for a spin-1/2 particle in \emph{any} external field. In particular, it is applicable for any stationary or nonstationary electromagnetic field. Taking it into account can be necessary at scattering. The problem of scattering is important for electromagnetic and strong interactions. However, it is not clear yet if strong interactions can be appropriately described by a relativistic Hamiltonian. Our analysis shows that the newly obtained correction is usually negligible in experiments with a particle moving in external electric and magnetic fields created by macroscopic devices. Certainly, it is also negligible for a particle in gravitational fields. Thus, the scattering theory is the main potential application of the result obtained. 

\section{Discussion and summary} \label{Discussion}

As a rule, interactions of a relativistic particle with external electric and magnetic fields can be appropriately described in the weak-field approximation. For this purpose, taking into account only terms linear in external fields and their derivatives is usually sufficient. For spin-1/2 particles, previous studies presented by Eqs. (\ref{eq12erk}) and (\ref{eq12A}) show that one can usually restrict oneself to taking into account a leading correction to Eq. (\ref{MHamD}).
For a particle moving in fields of laboratory devices, our newly obtained correction defined by the last term in Eq. (\ref{Hamwfin}) is too small 
even for contemporary high-precision experiments and can be neglected. However, the situation changes at scattering of a relativistic particle. In this case, the impact parameter $\mathfrak{b}$ can be of the order of the Compton wavelength $\lambdabar$ ($3.86\times 10^{-13}$ m for the electron and positron). When $p\sim mc$, we obtain that $\mathfrak{b}p\sim \hbar$ and the determined correction is of the same order as the precedent term in Eq. (\ref{eqpepnl}) and even ${\cal E}$. For scattering by light nuclei, in the considered case we obtain $|e|\Phi\sim\alpha cp\ll cp$. 
The presented estimates are made for the smallest distance between a scattered electron or positron and a nucleus.

The validity of the result obtained is confirmed by a comparison with the previously derived semi-relativistic FW Hamiltonian \cite{PRA2015,PRA2016,VJ,TMPFW}. We note that 4 terms in the relativistic Hamiltonian  (\ref{Hamwfin}) correspond to 11 ones in the semi-relativistic Hamiltonian (\ref{eq12erk}). Importantly, all the terms in the former Hamiltonian acquire a clear physical sense after replacing ${\cal E}$ and ${\cal O}$ with particle and field parameters. By contrast, the physical sense of numerous terms of the former Hamiltonian cannot be specified. The used approach is applicable for a determination of additional (quadratic in ${\cal E}$) terms in the \emph{relativistic} FW Hamiltonian. This task can be considered as an outlook. In NRQED, its solution would cover all nonzero terms up to the order of $m\alpha^8$ originating from the initial Dirac or Dirac-Pauli equation. In contrast with previously derived semi-relativistic FW Hamiltonians (which, in addition, differ from each other \cite{Zhou2023,Jentschura2024}), the relativistic FW Hamiltonian would have a clear physical sense. This advantage of \emph{analytically obtained} relativistic FW Hamiltonians could be important even for \emph{relativistic} quantum chemistry.

In summary, we used the exact operator of the FW transformation obtained by Eriksen \cite{E} and fulfilled the operator extraction of a square root in the Eriksen operator. Compared with the previously derived \cite{JMP} spin-dependent term, the obtained correction is characterized by an additional factor of the order of $\hbar/(rp)$, where $r\sim |\partial E_i/\partial x_j|/E$. This correction could be important for a description of scattering of a relativistic spin-1/2 particle. For this task, the use of a relativistic wave equation of the second order can be necessary. This equation with a correction similar to that to the FW Hamiltonian has been rigorously derived in the present study. A calculation of leading corrections to the FW Hamiltonian is a natural way for a development of the relativistic FW transformation method. The used approach can be applied for a derivation of corrections quadratic (bilinear) in ${\cal E}$ for a spin-1/2 particle in nonstationary external fields. 

\section*{Data availability statement}

All data that support the findings of this study are included within the article.




\begin{thebibliography}{99}

\bibitem{FW}
 L.\,L. Foldy, S.\,A. Wouthuysen, On the Dirac Theory of Spin 1/2
Particles and Its Non-Relativistic Limit,
Phys. Rev. \textbf{78}, 29 (1950).

\bibitem{NW}
T.\,D. Newton, E.\,P. Wigner,
Localized States for Elementary Systems,
Rev. Mod. Phys. \textbf{21}, 400 (1949).

\bibitem{Reply2019}
A. J. Silenko, Pengming Zhang, and Liping Zou, Silenko, Zhang, and Zou Reply, 
Phys. Rev. Lett. \textbf{122}, 159302 (2019).

\bibitem{PRAFW}
Liping Zou, Pengming Zhang, and A. J. Silenko, Position and spin in relativistic quantum mechanics,
Phys. Rev. A \textbf{101}, 032117 (2020). 

\bibitem{JMP}
A.\,J. Silenko, Foldy–Wouthuysen transformation for relativistic particles in external fields, J. Math. Phys. {\bf 44}, 2952 (2003).

\bibitem{JINRLett12}
A.\,J. Silenko, Classical limit of relativistic quantum mechanical equations in the Foldy-Wouthuysen representation,
Pis'ma Zh. Fiz. Elem. Chast. Atom. Yadra \textbf{10},
144 (2013) [Phys. Part. Nucl. Lett. \textbf{10}, 91 (2013)].

\bibitem{dVFor}
E. de Vries, Foldy--Wouthuysen Transformations and Related Problems,
Fortschr. Phys. \textbf{18}, 149 (1970).

\bibitem{CMcK}
J.\,P. Costella, B.\,H.\,J. McKellar,
The Foldy--Wouthuysen transformation,
Am. J. Phys. \textbf{63}, 1119 (1995).

\bibitem{E}
E. Eriksen, Foldy--Wouthuysen Transformation. Exact Solution with
Generalization to the Two-Particle Problem,
Phys. Rev. \textbf{111}, 1011 
(1958).

\bibitem{PRA2015}
A.\,J. Silenko, General method of the relativistic Foldy--Wouthuysen transformation
and proof of validity of the Foldy--Wouthuysen Hamiltonian,
Phys. Rev. A \textbf{91}, 022103 (2015).

\bibitem{PRAnonstat}
A. J. Silenko, Energy expectation values of a particle in nonstationary fields, 
Phys. Rev. A \textbf{91}, 012111 (2015).

\bibitem{BLP}
V. B. Berestetskii, E. M. Lifshitz, and L. P. Pitayevskii,
{\em Quantum Electrodynamics}, 2nd ed. (Pergamon, Oxford, 1982).

\bibitem{Pauli} W. Pauli, Relativistic Field Theories of Elementary Particles,
Rev. Mod. Phys. \textbf{13}, 203 (1941).

\bibitem{PRA2016}
A.\,J. Silenko, General properties of the Foldy-Wouthuysen transformation and applicability of the corrected original Foldy-Wouthuysen method, Phys. Rev. A \textbf{93}, 022108 (2016).

\bibitem{erik}
E. Eriksen and M. Korlsrud, Canonical Transformations of Dirac's Equation to Even Forms.
Expansion in Terms of the External Fields, Nuovo Cimento Suppl. \textbf{18},
1 (1960).

\bibitem{JMPcond}
V.\,P. Neznamov and A.\,J. Silenko, Foldy--Wouthuysen wave functions and conditions of transformation
between Dirac and Foldy--Wouthuysen representations, J. Math. Phys. \textbf{50},
122302 (2009).

\bibitem{PRAExpO}
A.\,J. Silenko, Exact form of the exponential Foldy-Wouthuysen transformation operator for
an arbitrary-spin particle, Phys. Rev. A \textbf{94}, 032104 (2016).

\bibitem{Valid} A.\,J. Silenko, Validation of the Eriksen Method
for the Exact Foldy--Wouthuysen Representation, Pis'ma Zh. Fiz. Elem. Chast. Atom. Yadra \textbf{10},
321 (2013) [Phys. Part. Nucl. Lett. \textbf{10}, 198 (2013)].

\bibitem{Steph}
S. Stephani, Zur Entwicklung der Dirac-Gleichung nach $c^{-2}$,
Ann. Phys. (Leipzig) {\bf 470}, 12 
(1965).

\bibitem{Reuse} F.\,A. Reuse, \textit{Electrodynamique et Optique Quantiques} (Presses Polytechniques et
Universitaires Romandes, Lausanne, 2007).

\bibitem{ultrafast}
Y. Hinschberger and P.-A. Hervieux, Foldy--Wouthuysen transformation applied to the interaction of an electron with
ultrafast electromagnetic fields,
\textit{Phys. Lett. A} \textbf{376}, 813 (2012).

\bibitem{FizElem} V. P. Neznamov, On the Theory of Interacting Fields in
Foldy-Wouthuysen Representation, Fiz. Elem. Chastits At. Yadra {\bf
37}, 152 (2006) [Phys. Part. Nucl. {\bf 37}, 86 (2006)].

\bibitem{PRA}
A.\,J. Silenko, Foldy-Wouthuysen transformation and semiclassical limit for relativistic particles in strong external fields, Phys. Rev. A \textbf{77}, 012116 (2008).

\bibitem{PengReiher}
D. Peng and M. Reiher, Exact decoupling of the relativistic Fock operator,
Theor. Chem. Acc. \textbf{131}, 1081 (2012).

\bibitem{relativistic}
E.\,I. Blount, Extension of the Foldy-Wouthuysen Transformation, Phys. Rev. \textbf{128}, 2454 (1962);
A.\,J. Silenko, Dirac equation in the Foldy-Wouthuysen representation describing the interaction of spin-1/2 relativistic particles with an external electromagnetic field, Theor. Math. Phys. {\bf 105}, 1224 
(1995);
A.\,J. Silenko, Foldy-Wouthuysen representation in the standard model of electroweak interactions, 
Theor. Math. Phys. {\bf 112}, 922 (1997); 
K.\,Y. Bliokh, Topological spin transport of a relativistic electron, Europhys. Lett. \textbf{72}, 7
(2005);
K.\,Y. Bliokh, On the Hamiltonian nature of semiclassical equations of motion in the presence of an electromagnetic field and Berry curvature, Phys. Lett. A \textbf{351}, 123 (2006);
P. Gosselin, A. B\'{e}rard, and H. Mohrbach, Semiclassical diagonalization of quantum Hamiltonian and equations of motion with Berry phase corrections, Eur. Phys. J. B \textbf{58},
137 
(2007);
P. Gosselin and H. Mohrbach, Diagonal representation for a generic matrix valued quantum Hamiltonian, Eur. Phys. J. C \textbf{64}, 495 
(2009).

\bibitem{DouglasKroll}
M. Douglas and N.\,M. Kroll, Quantum electrodynamical corrections to the fine structure of helium, Ann. Phys. (N.Y.) \textbf{82}, 89 (1974).

\bibitem{Hess}
B.\,A. Hess, Applicability of the no-pair equation with free-particle projection operators to atomic and molecular structure calculations, Phys. Rev. A \textbf{32}, 756 (1985);
Relativistic electronic-structure calculations employing a two-component no-pair formalism with external-field projection operators, \textbf{33}, 3742 (1986).

\bibitem{local}
D. Peng and M. Reiher, Local relativistic exact decoupling,
J. Chem. Phys. \textbf{136}, 244108 (2012).

\bibitem{QuantChem}
D. K\c{e}dziera and M. Barysz, Non-iterative approach to the infinite-order two-component (IOTC) relativistic theory and the non-symmetric algebraic Riccati equation, Chem. Phys. Lett. \textbf{446}, 176 (2007);
F. Aquilante \emph{et al}, MOLCAS 7: The Next Generation, J. Comput. Chem. \textbf{31}, 224 (2010);
D. Peng, N. Middendorf, F. Weigend, and M. Reiher, An efficient implementation of two-component relativistic exact-decoupling methods for large molecules, J. Chem. Phys. \textbf{138}, 184105 (2013).

\bibitem{ReiherWolf}
M. Reiher and A. Wolf, Exact decoupling of the Dirac Hamiltonian. I. General theory, J. Chem. Phys. \textbf{121}, 2037 (2004).

\bibitem{ReiherWolfNext}
M. Reiher and A. Wolf, Exact decoupling of the Dirac Hamiltonian. II. The generalized Douglas–Kroll–Hess transformation up to arbitrary order, J. Chem. Phys. \textbf{121}, 10945 (2004); A. Wolf and M. Reiher, Exact decoupling of the Dirac Hamiltonian. III. Molecular properties, J. Chem. Phys. \textbf{124}, 064102 (2006); Exact decoupling of the Dirac Hamiltonian. IV. Automated evaluation of molecular properties within the Douglas-Kroll-Hess theory up to arbitrary order, \textbf{124}, 064103 (2006).

\bibitem{Dyall} K.\,G. Dyall and K. Faegri, \textit{Introduction to relativistic quantum chemistry} (Oxford University Press, Oxford, 2007).

\bibitem{ReiherWolfBook}
M. Reiher and A. Wolf, \textit{Relativistic Quantum Chemistry: The Fundamental Theory of Molecular Science} (Wiley-VCH, Weinheim, 2009).

\bibitem{ChiouChen}
Dah-Wei Chiou and Tsung-Wei Chen, Exact Foldy-Wouthuysen transformation of the Dirac-Pauli Hamiltonian in the weak-field limit by the method of direct perturbation theory, Phys. Rev. A \textbf{94}, 052116 (2016).

\bibitem{RPJ}
A.\,J. Silenko, Quantum-mechanical description of the
electromagnetic interaction of relativistic particles with
electric and magnetic dipole moments, Russ. Phys. J. \textbf{48},
788 (2005). 

\bibitem{VJ}
E. de Vries, J.\,E. Jonker,
Non-relativistic approximations of the Dirac Hamitonian,
Nucl. Phys. B \textbf{6}, 213 
(1968).

\bibitem{TMPFW}
A.\,J. Silenko, Comparative analysis of direct and ``step-by-step''
Foldy--Wouthuysen transformation methods,
Teor. Mat. Fiz. \textbf{176}, 189 (2013)
[Theor. Math. Phys. \textbf{176}, 987 
(2013)].

\bibitem{Pachucki}
K. Pachucki, Higher-order effective Hamiltonian for light atomic systems, Phys. Rev. A \textbf{71}, 012503 (2005).

\bibitem{Pachucki2006}
K. Pachucki, $\alpha^4{\cal R}$ corrections to singlet states of helium, Phys. Rev. A \textbf{74}, 022512 (2006).

\bibitem{JentschuraCzarneckiPachucki}
U. D. Jentschura, A. Czarnecki, and K. Pachucki, Nonrelativistic QED approach to the Lamb shift, Phys. Rev. A \textbf{72}, 062102 (2005).

\bibitem{RHill}
R. J. Hill, G. Lee, G. Paz, and M. P. Solon, NRQED Lagrangian at order $1/\emph{M}^4$, Phys. Rev. D \textbf{87}, 053017 (2013).

\bibitem{PatkosPachucki}
V. Patk\'{o}\v{s}, V. A. Yerokhin, and K. Pachucki, Higher-order recoil corrections for triplet states of the helium atom,
Phys. Rev. A \textbf{94}, 052508 (2016).

\bibitem{Mei2014}
X. Mei, S. Zhao, and H. Qiao, Calculation of Higher-Order Foldy-Wouthuysen Transformation Hamiltonian, Chin. Phys. Lett. \textbf{31}, 063102 (2014).

\bibitem{Qiao2019}
W. Zhou, X. Mei and H. Qiao, Nonrelativistic quantum electrodynamic Hamiltonian and photon-exchange
interaction up to $m\alpha^8$ order, Phys. Rev. A \textbf{100}, 012513 (2019).

\bibitem{Zhou2023}
T. Chen, X. Mei, W. Zhou and H. Qiao, High-order Hamiltonian obtained by Foldy-Wouthuysen transformation up to the order
of $m\alpha^8$, Chin. Phys. B \textbf{32}, 083101 (2023).

\bibitem{ZhouJPhysB2023}
W. Zhou, X. Mei and H. Qiao, The $m\alpha^8$-order Foldy–Wouthuysen Hamiltonian and
relativistic corrections to Coulomb systems, J. Phys. B: At. Mol. Opt. Phys. \textbf{56}, 045001 (2023).

\bibitem{Jentschura2024}
U. D. Jentschura, Eighth-order Foldy-Wouthuysen transformation, Phys. Rev. A \textbf{110}, 012808 (2024).

\bibitem{CaswLep}
W. E. Caswell,  G. P. Lepage, Effective lagrangians for bound state problems in QED, QCD, and other field theories,
Phys. Lett. B \textbf{167}, 437 (1986).

\bibitem{exactFW}
A. G. Nikitin, On exact Foldy–Wouthuysen transformation, J. Phys. A: Math. Gen. \textbf{31}, 3297 (1998); B. Gon\c{c}alves, M. M. Dias J\'{u}nior and B. J. Ribeiro, Exact Foldy-Wouthuysen transformation for a Dirac theory with the complete set of \emph{CPT}-Lorentz invariance violating terms, Phys. Rev. D \textbf{90}, 085026 (2014); B. Gon\c{c}alves, B. J. Ribeiro, D. D. Pereira and M. M. Dias, The space–time torsion in the context of the exact Foldy–Wouthuysen transformation for a Dirac fermion, Int. J. Mod. Phys. A \textbf{31}, 1650075 (2016); B. Gon\c{c}alves, M. M. Dias J\'{u}nior and B. J. Ribeiro, Exact Foldy-Wouthuysen transformation for a Dirac theory revisited, Phys. Rev. D \textbf{99}, 096015 (2019); B. Gon\c{c}alves, M. M. Dias J\'{u}nior and L. F. Eleot\'{e}rio, Exact Foldy-Wouthuysen transformation for a Dirac equation describing the interaction of spin-1/2 relativistic particles with an external electromagnetic field, Int. J. Mod. Phys. A \textbf{38},  2350173 (2023).

\bibitem{PAN}
A. J. Silenko, Transformation of Operator Equations Describing the Interaction
of Relativistric Particles with an Electric Field, Yad. Fiz. {\bf 64}, 1048
(2001) [Phys. At. Nucl. {\bf 64}, 977 (2001)].

\bibitem{TMP2008}
A.J. Silenko, Hamilton operator and the semiclassical limit for
scalar particles in an electromagnetic field, Teor. Mat. Fiz. \textbf{156}, 398 (2008) [Theor.
Math. Phys. \textbf{156}, 1308 (2008)].

\bibitem{FV}
H. Feshbach and F. Villars, Elementary Relativistic Quantum Mechanics of Spin 0 and Spin 1/2 Particles, Rev. Mod. Phys. \textbf{30}, 24 (1958).

\bibitem{Dehmelt}
H. Dehmelt, Experiments with an isolated subatomic particle at rest, Rev. Mod. Phys. \textbf{62}, 
525 (1990).

\bibitem{Mooser}
A. Mooser, H. Kracke, K. Blaum, A. Bra\"{u}ninger, K. Franke, C. Leiteritz, W. Quint, C. C. Rodegheri, S. Ulmer, and J. Walz, Experiments with an isolated subatomic particle at rest, Phys. Rev. Lett. \textbf{110}, 140405 (2013).

\end{thebibliography}
\end{document}